\documentclass[12pt]{article}
\begin{document}

\title{\bf Teleparallel Energy-Momentum Distribution of Spatially
Homogeneous Rotating Spacetimes }

\author{M. Sharif \thanks{msharif@math.pu.edu.pk} and M. Jamil Amir
\thanks{mjamil.dgk@gmail.com}\\
Department of Mathematics, University of the Punjab,\\
Quaid-e-Azam Campus, Lahore-54590, Pakistan.}

\date{}

\maketitle

\begin{abstract}
The energy-momentum distribution of spatially homogeneous rotating
spacetimes in the context of teleparallel theory of gravity is
investigated. For this purpose, we use the teleparallel version of
Moller prescription. It is found that the components of
energy-momentum density are finite and well-defined but are
different from General Relativity. However, the energy-momentum
density components become the same in both theories under certain
assumptions. We also analyse these quantities for some special
solutions of the spatially homogeneous rotating spacetimes.
\end{abstract}

{\bf Keywords:} Teleparallel Theory, Energy.

\section{Introduction}
Till now, several attempts have been made to unify gravitation
with other interactions, including Einstein's, which led to the
investigation of basic structures of gravitation other than the
metric tensor. These structures yield the metric tensor as a by
product. Tetrad field is one of these structures which leads to
the theory of teleparallel gravity (TPG) [1,2]. TPG is an
alternative theory of gravity which corresponds to a gauge theory
of translation group [3,4] based on Weitzenb$\ddot{o}$ck geometry
[5]. This theory is characterized by the vanishing of curvature
identically while the torsion is taken to be non-zero. In TPG,
gravitation is attributed to torsion [4] which plays a role of
force [6]. In General Relativity (GR), gravitation geometrizes the
underlying spacetime. The translational gauge potentials appear as
a non-trivial part of the tetrad field and induce a teleparallel
(TP) structure on spacetime which is directly related to the
presence of a gravitational field. In some other theories [3-8],
torsion is only relevant when spins are important [9]. This point
of view indicates that torsion might represent additional degrees
of freedom as compared to curvature. As a result, some new physics
may be associated with it. Teleparallelism is naturally formulated
by gauging external (spacetime) translations which are closely
related to the group of general coordinate transformations
underlying GR. Thus the energy-momentum tensor represents the
matter source in the field equations of tetradic theories of
gravity like in GR.

The localization of energy (i.e., to express it as a unique tensor
quantity) has been a longstanding, open and controversial problem
in GR [10] which is still with out a definite answer. As a
pioneer, Einstein [11] introduced the energy-momentum
pseudo-tensor and then Landau-Lifshitz [12], Papapetrou [13],
Bergmann [14], Tolman [15] and Weinberg [16] proposed their own
prescriptions to resolve this issue. All these prescriptions can
provide meaningful results only in Cartesian coordinates. But
M$\ddot{o}$ller [17] introduced a coordinate-independent
prescription. The idea of coordinate-independent quasi-local mass
was introduced [18] by associating a Hamiltonian term to each
gravitational energy-momentum pseudo-tensor. Later, a Hamiltonian
approach in the frame of Schwinger condition [19] was developed,
followed by the construction of a Lagrangian density of TP
equivalent to GR [4,6,20,21]. Many authors explored many examples
in the framework of GR and found that different energy-momentum
complexes can give either the same [22] or different [23] results
for a given spacetime.

Mikhail et al. [24] defined the superpotential in the Moller's
tetrad theory which has been used to find the energy in TPG.
Vargas [25] defined the TP version of Bergman, Einstein and
Landau-Lifshitz prescriptions and found that the total energy of
the closed Friedman-Robinson-Walker universe is zero by using the
last two prescriptions. This agrees with the results of GR
available in literature [26]. Later, many authors [27] used TP
version of these prescriptions and showed that energy may be
localized in TPG. Keeping this point in mind, this paper is
addressed to investigate the energy-momentum distribution of
spatially homogeneous and rotating spacetimes by using the TP
version of M$\ddot{o}$ller prescription.

This paper is organized as follows. Section $2$ contains an
overview of the TP theory. In section $3$, we investigate the
energy-momentum distribution of spatially homogeneous rotating
spacetimes by using the TP version of M$\ddot{o}$ller
prescription. Section $4$ is devoted to discuss five special cases
of the metric representing spatially homogeneous rotating
spacetimes. The last section will furnish a summary and a
discussion of the results obtained.

\section{An Overview of the Teleparallel Theory}

The TP theory is based on Weitzenb$\ddot{o}$ck connection given as
[28]
\begin{eqnarray}
{\Gamma^\theta}_{\mu\nu}={{h_a}^\theta}\partial_\nu{h^a}_\mu,
\end{eqnarray}
where ${h_a}^\nu$ is a non-trivial tetrad. Its inverse field is
denoted by ${h^a}_\mu$ and satisfying the relations
\begin{eqnarray}
{h^a}_\mu{h_a}^\nu={\delta_\mu}^\nu; \quad\
{h^a}_\mu{h_b}^\mu={\delta^a}_b.
\end{eqnarray}
In this paper, the Latin alphabet $(a,b,c,...=0,1,2,3)$ will be used
to denote tangent space indices and the Greek alphabet
$(\mu,\nu,\rho,...=0,1,2,3)$ to denote spacetime indices. The
Riemannian metric in TP theory arises as a by product [4] of the
tetrad field given by
\begin{equation}
g_{\mu\nu}=\eta_{ab}{h^a}_\mu{h^b}_\nu,
\end{equation}
where $\eta_{ab}$ is the Minkowski metric
$\eta_{ab}=diag(+1,-1,-1,-1)$. For the Weitzenb$\ddot{o}$ck
spacetime, the torsion is defined as [2]
\begin{equation}
{T^\theta}_{\mu\nu}={\Gamma^\theta}_{\nu\mu}-{\Gamma^\theta}_{\mu\nu}
\end{equation}
which is antisymmetric w.r.t. its last two indices. Due to the
requirement of absolute parallelism, the curvature of the
Weitzenb$\ddot{o}$ck connection vanishes identically. The
Weitzenb$\ddot{o}$ck connection also satisfies the relation
\begin{equation}
{{\Gamma^{0}}^\theta}_{\mu\nu}={\Gamma^\theta}_{\mu\nu}
-{K^{\theta}}_{\mu\nu},
\end{equation}
where
\begin{equation}
{K^\theta}_{\mu\nu}=\frac{1}{2}[{{T_\mu}^\theta}_\nu+{{T_\nu}^
\theta}_\mu-{T^\theta}_{\mu\nu}]
\end{equation}
is the {\bf contortion tensor} and ${{\Gamma^{0}}^\theta}_{\mu\nu}
$ are the Christoffel symbols in GR.

Mikhail et al. [24] defined the super-potential of the
M$\ddot{o}$ller tetrad theory as
\begin{equation}
{U_\mu}^{\nu\beta}=\frac{\sqrt{-g}}{2\kappa}P_{\chi\rho\sigma}^
{\tau\nu\beta}[{V^\rho}g^{\sigma\chi} g_{\mu\tau}-\lambda
g_{\tau\mu} K^{\chi\rho\sigma}-(1-2\lambda)
g_{\tau\mu}K^{\sigma\rho\chi}],
\end{equation}
where
\begin{equation}
P_{\chi\rho\sigma}^{\tau\nu\beta}= {\delta_\chi}^{\tau}
g_{\rho\sigma}^{\nu\beta}+{\delta_\rho}^{\tau}
g_{\sigma\chi}^{\nu\beta}-{\delta_\sigma}^{\tau}
g_{\chi\rho}^{\nu\beta}
\end{equation}
and $ g_{\rho\sigma}^{\nu\beta}$ is a tensor quantity defined by
\begin{equation}
g_{\rho\sigma}^{\nu\beta}={\delta_\rho}^{\nu}{\delta_\sigma}^{\beta}-
{\delta_\sigma}^{\nu}{\delta_\rho}^{\beta}.
\end{equation}
$K^{\sigma\rho\chi}$ is the contortion tensor given by Eq.(6), $g$
is the determinant of the metric tensor $g_{\mu\nu}$, $\lambda$ is
the free dimensionless coupling constant of TPG, $\kappa$ is the
Einstein constant and $V_\mu$ is the basic vector field given by
\begin{equation}
{V_\mu}={T^\nu}_{\nu\mu}
\end{equation}
The energy-momentum density is defined as
\begin{equation}
\Xi_\mu^\nu= U_\mu^{\nu\rho},_\rho,
\end{equation}
where comma means ordinary differentiation. The momentum 4-vector
of M$\ddot{o}$ller prescription can be expressed as
\begin{equation}
P_\mu ={\int}_\Sigma {\Xi_\mu^0} dxdydz,
\end{equation}
where  $P_0$ gives the energy and $P_1$, $P_2$ and $P_3$  are the
momentum components while the integration is taken over the
hyper-surface element $\Sigma$ described by $x^0=t=constant$. The
energy may be given in the form of surface integral [17] as
\begin{equation}
E=\lim_{r \rightarrow \infty} {\int}_{{r=constant}}
{U_0}^{0\rho}u_\rho dS,
\end{equation}
where $u_\rho$ is the unit three-vector normal to the surface
element $dS$.

\section{Teleparallel Energy of Spatially Homogeneous Rotating Spacetimes }

After the pioneering work of Gamow [29] and G$\ddot{o}$del [30],
the idea of global rotation of the universe has become a favorite
topic in the research field of GR. The metric representing
spatially homogeneous universes with rotation but no shear can be
given as
\begin{equation}
ds^2=dt^2-dr^2-A(r)d\phi^2-dz^2+2B(r)dtd\phi,
\end{equation}
where $A(r)$ and $B=B(r)$ are arbitrary functions. The metric
given by Eq.(14) represents five spacetimes [30,31], which can be
achieved by choosing particular values of the metric functions $A$
and $B$. Using the procedure adopted in the papers [32,33], the
tetrad components of the above metric can written as
\begin{equation}
{h^a}_\mu=\left\lbrack\matrix { 1   &&&   0    &&&   B    &&&   0
\cr 0        &&& \cos\phi &&&   -\Delta\sin\phi    &&&   0 \cr 0
&&& \sin\phi&&&  \Delta\cos\phi&&& 0 \cr 0        &&&   0 &&& 0
&&& 1\cr } \right\rbrack
\end{equation}
with its inverse
\begin{equation}
{h_a}^\mu=\left\lbrack\matrix { 1   &&   0    && 0 && 0 \cr
\frac{B}{\Delta}\sin\phi && \cos\phi && -\frac{1}{\Delta}\sin\phi
&& 0 \cr- \frac{B}{\Delta}\cos\phi && \sin\phi &&
\frac{1}{\Delta}\cos\phi && 0 \cr 0 && 0 && 0 && 1\cr }
\right\rbrack.
\end{equation}
Here $\Delta=\Delta(r)=\sqrt{A+B^2}$. In view of Eqs.(15) and
(16), Eq.(1) yields the following non-vanishing components of the
Weitzenb$\ddot{o}$ck connection:
\begin{eqnarray}
{\Gamma^0}_{12}&=&-\frac{B}{\Delta}, \quad {\Gamma^0}_{21}
=B'-\frac {B\Delta'}{\Delta},\nonumber\\
{\Gamma^2}_{12}&=&\frac{1}{\Delta} ,\quad
{\Gamma^2}_{21}=\frac{\Delta'}{\Delta},\nonumber\\
{\Gamma^1}_{22}&=&-\Delta.
\end{eqnarray}
The corresponding non-vanishing components of the torsion tensor
are
\begin{eqnarray}
{T^0}_{12}&=&B'+\frac{B}{\Delta}(1-\Delta'),\nonumber\\
{T^2}_{12}&=&\frac{1}{\triangle}(\Delta'-1).
\end{eqnarray}
Substituting these values in Eq.(10) and then multiplying by
$g^{11}$ and $g^{33}$ respectively, we get
\begin{eqnarray}
V^1&=&\frac{1}{\Delta}(\Delta'-1),\\
V^3&=&0.
\end{eqnarray}
In view of Eqs.(18) and (6), the non-vanishing components of the
contorsion tensor are
\begin{eqnarray}
K^{100}&=&\frac{B}{\Delta^3}(B+B'\Delta-B\Delta')=-K^{010},\nonumber\\
K^{122}&=&\frac{1}{\Delta^3}
(\Delta'-1)=-K^{212},\nonumber\\
K^{102}&=&
K^{120}=\frac{B}{\Delta^3}(1-\Delta')+\frac{B'}{2\Delta^2}
=-K^{012}=-K^{210}, \nonumber\\
K^{021}&=&\frac{B'}{2\Delta^2}=-K^{201}.
\end{eqnarray}
It should be mentioned here that the contorsion tensor is
antisymmetric w.r.t. its first two indices. Making use of
Eqs.(19)-(21) in Eq.(7), we obtain the required independent
non-vanishing components of the supperpotential in
M$\ddot{o}$ller's tetrad theory as
\begin{eqnarray}
U_0^{01}&=&\frac{1}{2\kappa\Delta}\{(1+\lambda)BB'+2\Delta(1-\Delta')\}
=U_0^{10},\nonumber\\
U_2^{01}&=&\frac{1}{2\kappa\Delta}\{(1+\lambda)B^2B'+(1-\lambda)\Delta^2B'
+2B\Delta(1-\Delta')\}=-U_2^{10},\nonumber\\
U_0^{21}&=&-\frac{1}{2\kappa\Delta}(1+\lambda)B'=-U_0^{12}.
\end{eqnarray}
It is worth mentioning here that the supperpotential is skew
symmetric w.r.t. its last two indices. When we make use of
Eqs.(22) in Eq.(11), the energy density turns out to be
\begin{eqnarray}
\Xi_0^0&=&\frac{1}{2\kappa \Delta^2}\{(1+\lambda)(\Delta
BB''+\Delta B'^2-BB'\Delta')-2\Delta^2\Delta''\}.
\end{eqnarray}
For $\lambda=1$
\begin{equation}
{E^d}_{TPT}={E^d}_{GR}-\frac{\Delta''}{\kappa},
\end{equation}
where $E^d$ stands for energy density. It is clear that the energy
density in both the theories is same in the case, if possible,
$\Delta'=constant$. The only non-zero component of momentum
density is along $\phi$ direction and (for $\lambda=1$) is given
by
\begin{eqnarray}
\Xi_2^0=\frac{1}{\kappa
\Delta^2}\{\Delta(AB''-A''B)-\Delta'(AB'-A'B)\}-\frac{1}{\kappa}(B''\Delta-2B'\Delta').
\end{eqnarray}
that is,
\begin{equation}
{M^d}_{TPT}={M^d}_{GR}-\frac{1}{\kappa}(B''\Delta-2B'\Delta'),
\end{equation}
where $M^d$ stands for momentum density. Again in the case, if
possible, when $B$ is constant the momentum density in TPT and GR
become same.

\section{Some Special Cases of Spatially Homogeneous Rotating Spacetimes}

In this section, we would like to mention some special cases of
spatially homogeneous rotating spacetimes. At the end, the
results for these cases will be given in tables.\\
\textbf{\textit{1. The Reboucas(Rb) Spacetime}}  [31]\\
By choosing $A(r)=-(1+3{c^*}^2)$ and $B(r)=2c^*$, where
$c^*=cosh2r$ and $s^*=sinh2r$, the metric (14) takes the form
\begin{equation}
ds^2=dt^2-dr^2+(1+3{c^*}^2)d\phi^2-dz^2+4c^*dtd\phi,
\end{equation}
\textit{\textbf{2. The Som-Raychaudhuri(SR) Spacetime}} [31]\\
By choosing $A(r)=r^2(1-r^2)$ and $B(r)=r^2$, the metric (14)
reduces to
\begin{equation}
ds^2=dt^2-dr^2-r^2(1-r^2)d\phi^2-dz^2+2r^2dtd\phi,
\end{equation}
\textit{\textbf{3. The Hoenselaers-Vishveshwara(HV) Spacetime}} [31]\\
By choosing $A(r)=-\frac{1}{2}(c-1)(c-3)$ and $B(r)=c-1$, where
$c=coshr$, the metric (14) implies that
\begin{equation}
ds^2=dt^2-dr^2+\frac{1}{2}(c-1)(c-3)d\phi^2-dz^2+2(c-1)dtd\phi,
\end{equation}
\textit{\textbf{4. The G$\ddot{o}$del-Friedmann(GF) Spacetime}} [31]\\
By choosing $A(r)=s^2(1-s^2)$ and $B(r)=\sqrt{2}s^2$, where
$s=sinhr$, the metric (14) turns out as
\begin{equation}
ds^2=dt^2-dr^2-s^2(1-s^2)d\phi^2-dz^2+2\sqrt{2}s^2dtd\phi,
\end{equation}
\textit{\textbf{5. The Stationary G$\ddot{o}$del(SG) Spacetime}}
[30]\\
By choosing $A(r)=-\frac{1}{2}e^{2ar}$ and $B(r)=e^{ar}$, where
$a=constant$, the metric (14) becomes
\begin{equation}
ds^2=dt^2-dr^2+\frac{1}{2}e^{2ar}d\phi^2-dz^2+2e^{ar}dtd\phi,
\end{equation}
The corresponding energy-momentum densities of these spacetimes
are given in the following tables:\\

{\bf {\small Table 1:}} {\small \textbf{Energy Densities of the
Reduced Cases of Spatially Homogeneous Rotating Spacetimes}}

\vspace{0.5cm}

\begin{center}
\begin{tabular}{|c|c|c|c|}
\hline{\bf Spacetime}&{\bf In GR }[38]&
{\bf In TPT}& {\bf Remarks} \\
\hline Rb & $\Theta_0^0=\frac{2s^*}{\pi}$ &
$\Xi_0^0=\frac{3s^*}{2\pi}$
 & $\Xi_0^0=\frac{3}{4}\Theta_0^0$\\
\hline SR & $\Theta_0^0=\frac{r}{2\pi}$ & $\Xi_0^0=\frac{r}{2\pi}$
 & $\Xi_0^0=\Theta_0^0$ \\
\hline  HV & $\Theta_0^0=\frac{s}{4\sqrt{2}\pi}$ &
$\Xi_0^0=\frac{s}{8\sqrt{2}\pi}$
 & $\Xi_0^0=\frac{1}{2}\Theta_0^0$ \\
\hline  GF & $\Theta_0^0=\frac{cs}{\pi}$ &
$\Xi_0^0=\frac{cs}{2\pi}$
 & $\Xi_0^0=\frac{1}{2}\Theta_0^0$ \\
\hline  SG & $\Theta_0^0=\frac{a^2}{4\sqrt{2}\pi}e^{ar}$ &
$\Xi_0^0=\frac{a^2}{8\pi}e^{ar}$
 & $\Xi_0^0=\frac{1}{\sqrt{2}}\Theta_0^0$  \\
\hline
\end{tabular}
\end{center}

{\bf {\small Table 2:}} {\small \textbf{Momentum Densities of the
Reduced Cases of Spatially Homogeneous Rotating Spacetimes}}

\vspace{0.5cm}

\begin{center}
\begin{tabular}{|c|c|c|c|}
\hline{\bf Spacetime}&{\bf In GR} [38]&
{\bf In TPT}& {\bf Remarks} \\
\hline Rb & $\Theta_2^0=\frac{6c^*s^*}{\pi}$ &
$\Xi_2^0=\frac{5s^*c^*}{\pi}$
 & $\Xi_2^0=\frac{5}{6}\Theta_2^0$ \\
\hline  SR & $\Theta_2^0=\frac{r^3}{\pi}$ &
$\Xi_2^0=\frac{4r^3+r}{4\pi}$
 & $\Xi_2^0=\Theta_2^0+\frac{r}{4\pi}$ \\
\hline  HV & $\Theta_2^0=\frac{s(c-1)}{4\sqrt{2}\pi}$ &
$\Xi_2^0=\frac{s(3c-2)}{8\sqrt{2}\pi}$
 & $\Xi_2^0=\Theta_2^0+\frac{cs}{8\sqrt{2}\pi}$ \\
\hline  GF & $\Theta_2^0=\frac{\sqrt{2}cs^3}{\pi}$ &
$\Xi_2^0=\frac{cs(5s^2+c^2)}{2\sqrt{2}\pi}$
 & $\Xi_2^0=\frac{5}{4}\Theta_2^0+\frac{c^3s}{2\sqrt{2}\pi}$ \\
\hline SG & $\Theta_0^0=\frac{a^2}{4\sqrt{2}\pi}e^{2ar}$ &
$\Xi_2^0=\frac{3a^2}{8\sqrt{2}\pi}e^{2ar}$ & $\Xi_2^0=\frac{3}{2}\Theta_2^0$ \\
\hline
\end{tabular}
\end{center}

\section{Summary and Discussion}

There is a large literature available [34] about the study of TP
versions of the exact solutions of GR. Recently, Pereira, et al.
[32] obtained the TP versions of the Schwarzschild and the
stationary axisymmetric Kerr solutions of GR. They proved that the
axial-vector torsion plays the role of the gravitomagnetic
component of the gravitational field in the case of slow rotation
and weak field approximations. In previous papers [33,35-37], we
have found the TP versions of the Friedmann models,
Lewis-Papapetrou spacetimes, stationary axisymmetric
Einstein-Maxwell solutions and also discussed the energy-momentum
distribution in last three papers.

The problem of localization of energy has been reconsidered, in
the frame work of TPG, by many scientists. Some authors [25,27]
showed that energy-momentum can also be localized in this theory.
It has been shown that the results of the two theories are either
same or disagree with each other for a given spacetime.
M$\ddot{o}$ller showed that a tetrad description of a
gravitational field equation allows a more satisfactory treatment
of the energy-momentum complex than does GR.

Currently [35-37], we considered some particular spacetimes and
calculated the energy-momentum densities and found that the
results disagree in general but can coincide under certain
conditions. In this paper, we extended the work and explore the
energy-momentum distribution of spatially homogeneous rotating
spacetimes by using the TP version of M$\ddot{o}$ller's
prescription. We found that the results are not generally same as
found in the context of GR [38]. If, possibly, we choose
$\Delta'$=costant(as in the case of SR spacetime), then
\begin{equation}
{E^d}_{TPG}={E^d}_{GR}
\end{equation}
and for $B$=constant
\begin{equation}
{M^d}_{TPG}={M^d}_{GR}
\end{equation}
It should be worth mentioning here that only the non-vanishing
component of the momentum density is along $\phi$-direction which
is due the cross term $dtd\phi$ involving in the metric (14). It
is similar to the case of stationary axisymmetric Einstein-Maxwell
solutuons [35]. Further, we consider some special cases and found
their energy-momentum distribution by replacing the corresponding
values of the metric functions. It is shown that the energy
density is same in both the theories only in the case of SR
spacetime while for the other cases it turns out as multiple of
some real number, as given in table 1. The relation between the
components of momentum density in GR [38] and TPT are given in
last column of table 2 for every reduced spacetime.

\vspace{0.5cm}

{\bf Acknowledgment}

\vspace{0.5cm}

We would like to thank Higher Education Commission Islamabad,
Pakistan for its financial support during this work.
\vspace{1.5cm}


{\bf References}

\begin{description}

\item{[1]} M$\ddot{u}$ller-Hoisson, F. and Nitsch, J.: Phys. Rev. {\bf D28}
           (1983)718.

\item{[2]} De Andrade, V. C. and Pereira, J.G.: Gen. Rel. Grav. {\bf 30}(1998)263.

\item{[3]} Hehl, F.W., McCrea, J.D., Mielke, E.W. and Ne'emann, Y.: Phys.
           Rep. {\bf 258}(1995)1.

\item{[4]} Hayashi, K. and Shirafuji, T.: Phys. Rev. {\bf D19}(1979)3524.

\item{[5]} Weitzenb$\ddot{o}$ck, R.: {\it Invarianten Theorie}
           (Gronningen: Noordhoft, 1923).

\item{[6]} De Andrade, V.C. and Pereira,  J.G.: Phys. Rev. {\bf
           D56}(1997)4689.

\item{[7]} Blagojecvic, M. {\it Gravitation and Gauge Symmetries} (IOP
           publishing, 2002).

\item{[8]} Hammond, R.T.: Rep. Prog. Phys. {\bf 65}(2002)599.

\item{[9]} Gronwald, F. and Hehl, F.W.: {\it On the Gauge Aspects of Gravity,
           Proceedings of the 14th School of Cosmology and Gravitation},
           Eric, Italy ed. Bergmann, P.G. et al. (World Scientific, 1996).

\item{[10]} Misner, C.W., Thorne, K.S. and Wheeler, J.A.: \textit{Gravitation}
            (Freeman, New York, 1973).

\item{[11]} Einstein, A.: Sitzungsber. Preus. Akad. Wiss. Berlin (Math. Phys.)
            778(1915), Addendum \textit{ibid} 779(1915).

\item{[12]} Landau, L.D. and Lifshitz, E.M.: \textit{The Classical Theory
            of Fields} (Addison-Wesley Press, New York, 1962).

\item{[13]} Papapetrou, A.: \textit{Proc. R. Irish Acad. } \textbf{A52}(1948)11.

\item{[14]} Bergman, P.G. and Thomson, R.: Phys. Rev. \textbf{89}(1958)400.

\item{[15]} Tolman, R.C.: \textit{Relativity, Thermodynamics and
           Cosmology} (Oxford University Press, Oxford, 1934).

\item{[16]} Weinberg, S.: \textit{Gravitation and Cosmology} (Wiley, New
           York, 1972).

\item{[17]} M$\ddot{o}$ller, C.: Ann. Phys. (N.Y.) \textbf{4}(1958)347.

\item{[18]} Chang, C.C. and Nester, J.M.: Phys. Rev. Lett. \textbf{83}
            (1999)1897 and references therein.

\item{[19]} Schwinger, J.: Phys. Rev. \textbf{130}(1963)1253.

\item{[20]} De Andrade, V.L, Guillen, L.C.T and Pereira, J.G.: Phys. Rev. Lett.
            {\bf 84}(2000)4533.

\item{[21]} Aldrovendi, R. and Pereira, J.G.: {\it An Introduction to
            Gravitation Theory} (preprint).

\item{[22]} Virbhadra, K.S.: Phys. Rev. \textbf{D60}(1999)104041;
            \textit{ibid} \textbf{D42}(1990)2919; Phys. Lett.
            \textbf{B331}(1994)302;
            Virbhadra, K.S. and Parikh, J.C.: Phys. Lett. \textbf{B317}(1993)312;
            Rosen, N. and Virbhadra, K.S.: Gen. Rel. Grav. \textbf{25}(1993)429;
            Xulu, S.S.: Astrophys. Space Sci. \textbf{283}(2003)23.

\item{[23]} Sharif, M.: Int. J. Mod. Phys. \textbf{A17}(2002)1175;
            \textit{ibid} \textbf{A18}(2003)4361; \textbf{A19}(2004)1495;
            \textbf{D13}(2004)1019;
            Sharif, M. and Fatima, T.: Nouvo Cim; \textbf{B120}(2005)533;
            Int. J. Mod. Phys. \textbf{A20}(2005)4309.

\item{[24]} Mikhail, F.I., Wanas, M.I., Hindawi, A. and Lashin, E.I.: Int. J. Theo.
            Phys. \textbf{32}(1993)1627.

\item{[25]} Vargas, T.: Gen. Rel. Grav. \textbf{36}(2004)1255.

\item{[26]} Penrose, R.: \textit{Proc. Roy. Soc., London} \textbf{A381}(1982)53;
            Tod, K.P.: \textit{Proc. Roy. Soc., London} \textbf{A388}(1983)457;
            Rosen, N.: Gen. Rel. Grav. \textbf{26}(1994)323.

\item{[27]} Nashed, G.G.L.: Nuovo Cim. \textbf{B119}(2004)967;
            Salti, M., Havare, A.: Int. J. Mod. Phys.
            \textbf{A20}(2005)2169;
            Salti, M.: Int. J. Mod. Phys. \textbf{A20}(2005)2175; Astrophys. Space Sci.
            \textbf{229}(2005)159;
            Aydogdu, O. and Salti, M.: Astrophys. Space Sci.
            \textbf{229}(2005)227;
            Aydogdu, O., Salti, M. and Korunur, M.: Acta Phys. Slov.
            \textbf{55}(2005)537; Sezgin, A., Melis, A. and Tarhan, I.:
            Acta Physica Polonica \textbf{B} (to appaer).

\item{[28]} De Andrade, V.L, Guillen, L.C.T and Pereira, J.G.:\emph{ An Introduction to Geometrical Physics}
            (World Scientific)

\item{[29]} Gamow, G.: Nature (London) \textbf{158}(1946)549.

\item{[30]} G$\ddot{o}$del, K.: Rev. Mod. Phys. \textbf{21}(1949)447.

\item{[31]} Krori, K.D., Borgohain, P. and Das, D.(Kar): J. Math.
            Phys. \textbf{29}(1988)1645.

\item{[32]} Pereira, J.G., Vargas, T. and Zhang, C.M.: Class. Quantum Grav.
            {\bf 18}(2001)833.

\item{[33]} Sharif, M. and Amir, M. Jamil.: Gen. Rel. Grav. {\bf 38}(2006)1735.

\item{[34]}  Hehl, F.W. and Macias, A.: Int. J. Mod. Phys. {\bf D8}(1999)399;
            Obukhov, Yu N., Vlachynsky, E.J., Esser, W., Tresguerres, R. and
            Hehl, F.W.: Phys. Lett. {\bf A220}(1996)1;
            Baekler, P., Gurses, M., Hehl, F.W. and McCrea, J.D.: Phys. Lett.
            {\bf A128}(1988)245;
            Vlachynsky, E.J. Esser, W., Tresguerres, R. and Hehl, F.W.: Class.
            Quantum Grav. {\bf 13}(1996)3253;
            Ho, J.K., Chern, D.C. and Nester, J.M.: Chin. J. Phys. {\bf
            35}(1997)640;
            Hehl, F.W., Lord, E.A. and Smally, L.L.: Gen. Rel. Grav. {\bf 13}
            (1981)1037;
            Kawa, T. and Toma, N.: Prog. Theor. Phys. {\bf 87}(1992)583;
            Nashed, G.G.L.: Phys. Rev. \textbf{D66}(2002)060415; Gen. Rel.
            Grav. \textbf{34}(2002)1074.

\item{[35]} Sharif, M. and Amir, M. Jamil.: Gen. Rel. Grav. {\bf 39}(2007)989.

\item{[36]} Sharif, M. and Amir, M. Jamil.: Mod. Phys. Lett. \textbf{A22}(2007)425.

\item{[37]} Sharif, M. and Amir, M. Jamil.: Mod. Phys. Lett. \textbf{A}(2007, to appear).

\item{[38]} Havare, A., Salti, M. and Yetkin, T.: \emph{ On the Energy-Momentum
            Densities of the Cylindrically Symmetric Gravitational Waves}: gr-qc/0502057.
\end{description}
\end{document}